\documentclass[twocolumn,prl,amsmath,amssymb,floatfix,10pt,showpacs]{revtex4}
\usepackage{SIunits,graphicx,psfrag,upgreek}

\usepackage{times}
\newcommand\secc[1]{\vskip0.5pc\noindent\textit{\textbf{#1}}.}

\newcommand\sfrac[2]{\leavevmode
  \raise.5ex\hbox{\tiny #1}\kern-.17em
  \raise.16ex\hbox{$\scriptscriptstyle/$}\kern-.16em
  \lower.25ex\hbox{\tiny #2}}
\newcommand\half{{\ensuremath\sfrac12}}

\let\tilde\widetilde
\let\bar\overline

\DeclareMathAlphabet{\mathitbf}{OML}{cmm}{b}{it}
\DeclareMathAlphabet{\mathit}{OML}{cmm}{m}{it}
\let\pi\uppi
\newcommand\uth{\uptheta\mkern-.5mu}
\newcommand\ur{\mathrm{r}}

\renewcommand\vec{\ensuremath\mathitbf}
\newcommand\vx{\vec x}
\newcommand\vxi{x}
\newcommand\vxt{\tilde{\vec x}}
\newcommand\vz{\boldsymbol\zeta}

\newcommand\vzt{\tilde{\boldsymbol\zeta}}
\newcommand\vzti{\tilde\zeta}
\newcommand\vf{\vec f}
\newcommand\vv{\vec v}
\newcommand\vth{\boldsymbol\theta}
\renewcommand\vr{\vec r}
\newcommand\D{\mathbf{D}}
\newcommand\K{\mathbf{K}}
\newcommand\Ga{\mathbf{\Gamma}}
\newcommand\R{\mathbf{R}}
\newcommand\Gco{\mathbf{G}}
\newcommand\W{\mathbf{W}}
\newcommand\Om{\mathbf{\Omega}}

\newcommand\kB{\ensuremath k_\mathrm{B}}
\newcommand\eqmin{\ensuremath_{*}}

\newcommand\ud{\mathop{}\!\mathrm{d}}
\newcommand\eff{{\textrm{(eff)}}}
\newcommand\run{{\textrm{run}}}
\newcommand\str{{\mathrm{stretch}}}
\newcommand\orient{{\mathrm{orient}}}
\newcommand\CTD{\ensuremath_{\mathrm{CTD}}}
\newcommand\NTD{\ensuremath_{\mathrm{NTD}}}
\newcommand\rot{\ensuremath^{\mathrm{(rot)}}}
\newcommand\trans{\ensuremath^{\mathrm{(tr)}}}
\newcommand\poise{\ensuremath\mathrm{P}}

\makeatletter
\newcommand\unspace{\futurelet\@foo\@unspace}
\newcommand\@unspace{\ifx\@foo\par
  \expandafter\@gobble\else\expandafter\g@bblesp@ce\fi}
\newcommand\g@bblesp@ce[1]{#1}
\makeatother

\newcommand\eqr[1]{(\ref{#1})}

\begin{document}

\title{Coarse-grained protein-protein stiffnesses and dynamics
  from all-atom simulations}
\author{Stephen D. Hicks}\email{sdh33@cornell.edu}
\author{C. L. Henley}\email{clh@ccmr.cornell.edu}
\affiliation{%
  Laboratory of Atomic and Solid State Physics,
  Cornell University,
  Ithaca, NY 14853-2501}

\begin{abstract}
Large protein assemblies, such as virus capsids, may be coarse-grained
as a set of rigid domains linked by generalized (rotational and
stretching) harmonic springs.  We present a method to obtain the
elastic parameters and overdamped dynamics for these springs from
all-atom molecular dynamics simulations of one pair of domains at a
time.  The computed relaxation times of this pair give a consistency
check for the simulation, and (using a fluctuation-dissipation
relationship) we find the corrective force needed to null systematic
drifts.  As a first application we predict the stiffness of an HIV
capsid layer and the relaxation time for its breathing mode.
\end{abstract}

\pacs{87.10.Pq,87.15.ap,87.15.hg,87.15.Ya}

\maketitle

Large protein assemblies---our particular interest is the capsids
(shells) of viruses---are pertinent to most of the soft-matter
physics in cells; how can one calculate their elastic properties and
corresponding dynamics?  Such assemblies are too large to handle by
all-atom simulations, but numerical coarse graining techniques are
opening the door to direct simulations~\cite{arkhipov2006sdv}.
Nevertheless, we still prefer simplified parametrizations for the
purposes of human understanding, analytic treatment, transmission to
other researchers, and building up coarse-grained
models~\cite{bib:models}.  In this paper we propose an approach to
extract these simplified parameters from all-atom molecular dynamics
(MD) of small subsystems.

Assume we already know how to extract an appropriate subsystem and an
intelligent way to project the $3N$ coordinates of its configurations
onto s small number $\vx$.  (Below, we do this explicitly for the case
of the HIV capsid protein.)  Our aim is, from the observed
trajectories $\vx(t)$, to extract the parameters for an effective
Hamiltonian and equation of motion.  We will model $\vx(t)$ as an
overdamped random walk in a biased harmonic potential.  This walk is
parametrized primarily by two important tensors: one to describe the
shape of the harmonic well, and the other to describe the (mainly
hydrodynamic) damping and the associated stochastic noise.  Combining
these tensors gives a matrix whose eigenvalues are the relaxation
rates.  With detailed measurement of the dynamics, we can identify
whether the simulation is equilibrated during the simulation time, and
can compute the external forces we must add so as to measure the
behavior near the biologically proper configuration.  This is similar
in spirit to computing a potential of mean force or free energy
landscape with Jarzynski's equality\cite{jarzynski}, except that our
coarse-grained $\vx$ has more than one component, and (at minimum)
represents angular degrees of freedom in addition to stretching.  As
an application, we simulate the important inter-domain interactions in
the HIV capsid and estimate the Young's modulus and Poisson ratio of
the capsid lattice, as well as the relaxation rate of the breathing
mode.

\secc{Coarse graining as stochastic dynamics} We represent our system
as a vector of generalized coordinates $\vxi_i$, $i=1\ldots N$, where
$N$ is far smaller than the number of atoms and is obtained by some
form of coarse-graining.  Our objective is to parametrize and
determine from simulation (i) an effective free energy potential
function $U(\vx)$, and (ii) an equation of motion, for the
coarse-grained coordinates.

We assume the coarse-grained degrees of freedom are overdamped:
this is true at time scales much longer than the ``ballistic scale''
of local bond vibrations ($t_\mathrm{bal}\sim1\pico\second$).
Then the  dynamics is a continuous-time random walk:
\begin{equation}\label{eq:eom}
  \frac{\ud\vx}{\ud t} = \Ga\vf(\vx,t) + \vz(t),
\end{equation}
where $\Ga$ is the (symmetric) {\it mobility tensor}, $\vf(\vx,t)$ is
the force, $\vz(t)$ is a (Gaussian) stochastic function satisfying
\begin{equation}
  \left\langle\vz(t)\otimes\vz(t')\right\rangle = 2\D\delta(t-t'),
\end{equation}
and $\D$ is the {\it diffusion tensor}.
For detailed balance, $\D=\kB T\Ga$ at temperature $T$.
We can expand the potential to second order about a point $\vx\eqmin$,
\begin{equation} \label{eq:pot}
  U(\vx) = U_0 - \vf\eqmin\cdot (\vx-\vx\eqmin) +
    \frac{1}{2} (\vx-\vx\eqmin) \K (\vx-\vx\eqmin),
\end{equation}
where $\K$ is the (symmetric) \emph{stiffness tensor}; then the force
in~\eqr{eq:eom} is $\vf(\vx) = \vf\eqmin-\K(\vx-\vx\eqmin)$.  From
measuring coordinate covariances in the simulation, we obtain $\K$:
\begin{equation}
\label{eq:Kmeasure}
   \Gco \equiv  
   \bigl\langle [\vx-\vx\eqmin] \otimes [\vx-\vx\eqmin] \bigr\rangle
   = \kB T\K^{-1}.
\end{equation}

If the static effective potential were our only interest, and if our
runs were always long enough to equilibrate our system, there would
have been no need to model the dynamics~\eqr{eq:eom}.  As we do need
the dynamics, we determine the diffusion tensor $\D$ (and hence $\Ga$)
by measuring the correlation function at short times between the
ballistic and relaxation times scales (see below) during which the
deterministic term in~\eqr{eq:eom} is less important than the noise:
\begin{equation}
  \label{eq:Dmeasure}
    \D = \frac{\bigl\langle
                [\vx(t')-\vx(t)] \otimes [\vx(t')- \vx(t)]
         \bigr\rangle}{2|t'-t|}
    \equiv \frac{ \W(t'-t)}{2|t'-t|}.
\end{equation}
We average $\W(\Delta t)/|\Delta t|$ over possible offsets $\Delta
t\gg t_\mathrm{bal}$, inversely weighted by the expected variances
$\propto (\Delta t)^3$.  This weighting also ensures our estimate has
negligible contribution from $t$ comparable to the relaxation times,
at which times $\W(t)$ is no longer linear in $t$.  Notice that since
$\Ga$ pertains to short-time dynamics, it is correctly measured even
in runs too short to equilibrate in the potential well.

If we transform into coordinates $\vxt \equiv \Ga^{-\half}\vx$ then the
equation of motion becomes
\begin{equation}\label{eq:eomt}
  \frac{\ud\vxt}{\ud t}
    = \Ga^\half\vf\eqmin -\R\bigl(\vxt-\vxt\eqmin\bigr) + \vzt(t),
\end{equation}
where
\begin{equation}\label{eq:noiset}
  \bigl\langle\vzti_\alpha(t)\vzti_\beta(t')\bigr\rangle
    = 2\kB T\delta_{\alpha\beta}\delta(t-t'),
\end{equation}
and the {\it relaxation matrix\/} $\R=\Ga^\half\K\Ga^\half$ (which has
units [time]$^{-1}$) is simply the stiffness tensor in our transformed
frame.  The eigenvalues of $\R$ are the decay rates $\tau_\alpha^{-1}$
for the relaxation normal modes $\alpha$.

The correlation time for a mode is the same as its relaxation time, so
the relative error in $\K$ for mode $\alpha$ is of order
$\sqrt{\tau_\alpha/\tau_\run}$, where $\tau_\run$ is the total run
time.  Thus, if all the $\tau_\alpha \ll \tau_\run$, our
estimate~\eqr{eq:Kmeasure} of $\K$ is valid.  But if $\tau_\alpha \sim
\tau_\run$ for some direction, not only are errors large, but the
initial deviation may still be relaxing over the entire run, which is
often visible as a steady drift of the coordinates with mean velocity
$\bar\vv$.  Averaging over time gives a large spurious variance in the
drifting directions, leading to an underestimate of the corresponding
stiffness.

\begin{figure*}%
  \psfrag{time}[tc]{time (ns)}%
  \psfrag{pos}[bc]{mode coordinate ($\sqrt{\mathrm{ns}}$)}%
  \hbox to \textwidth{\strut\hfill
   \vtop{\hsize=0.24\textwidth
    \hbox to \hsize{\includegraphics[width=\hsize,clip]{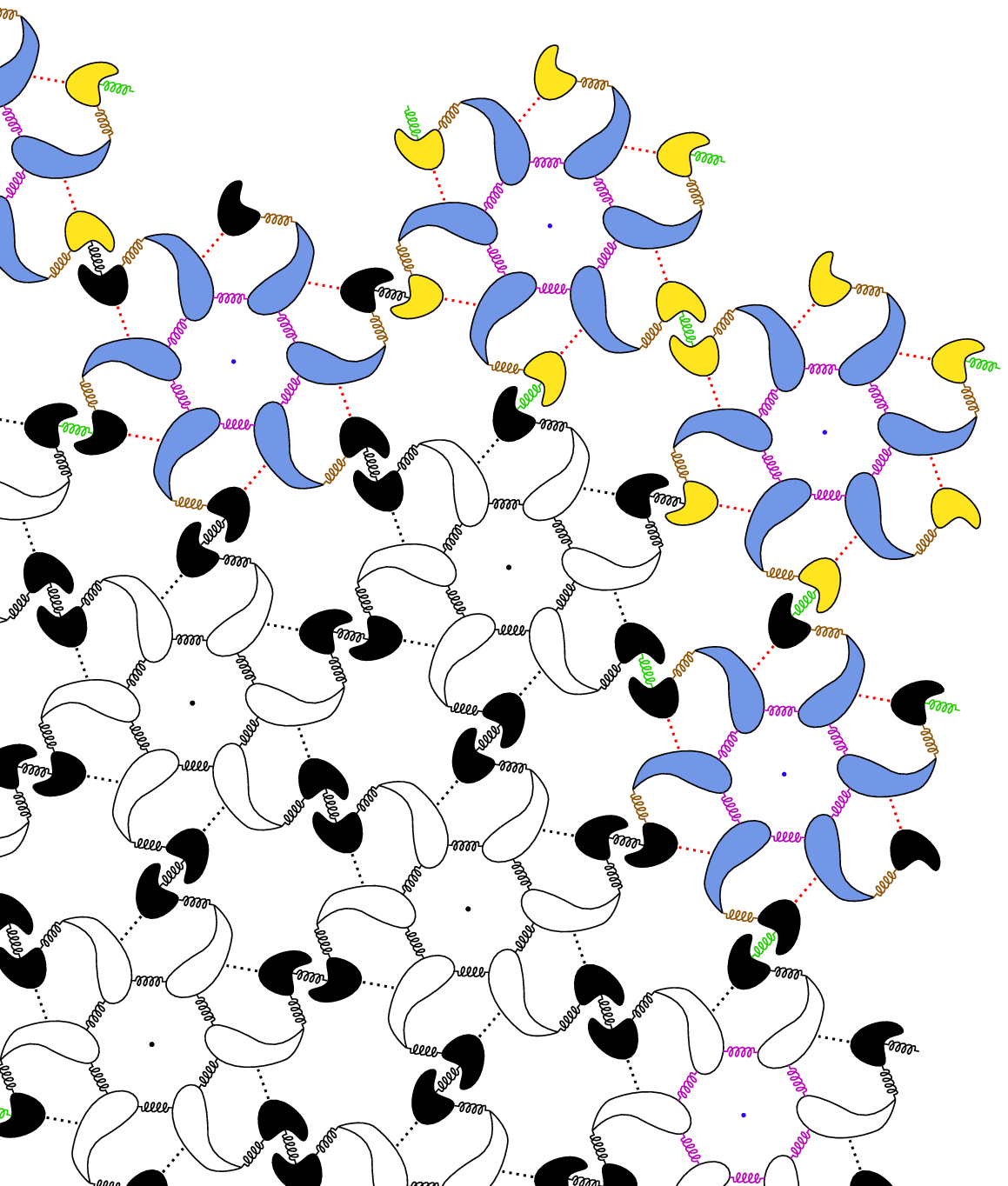}}
    \hbox to \hsize{\hfill(a)\hfill}}\hfill
   \vtop{\hsize=0.36\textwidth
    \hbox to \hsize{\includegraphics[width=\hsize]{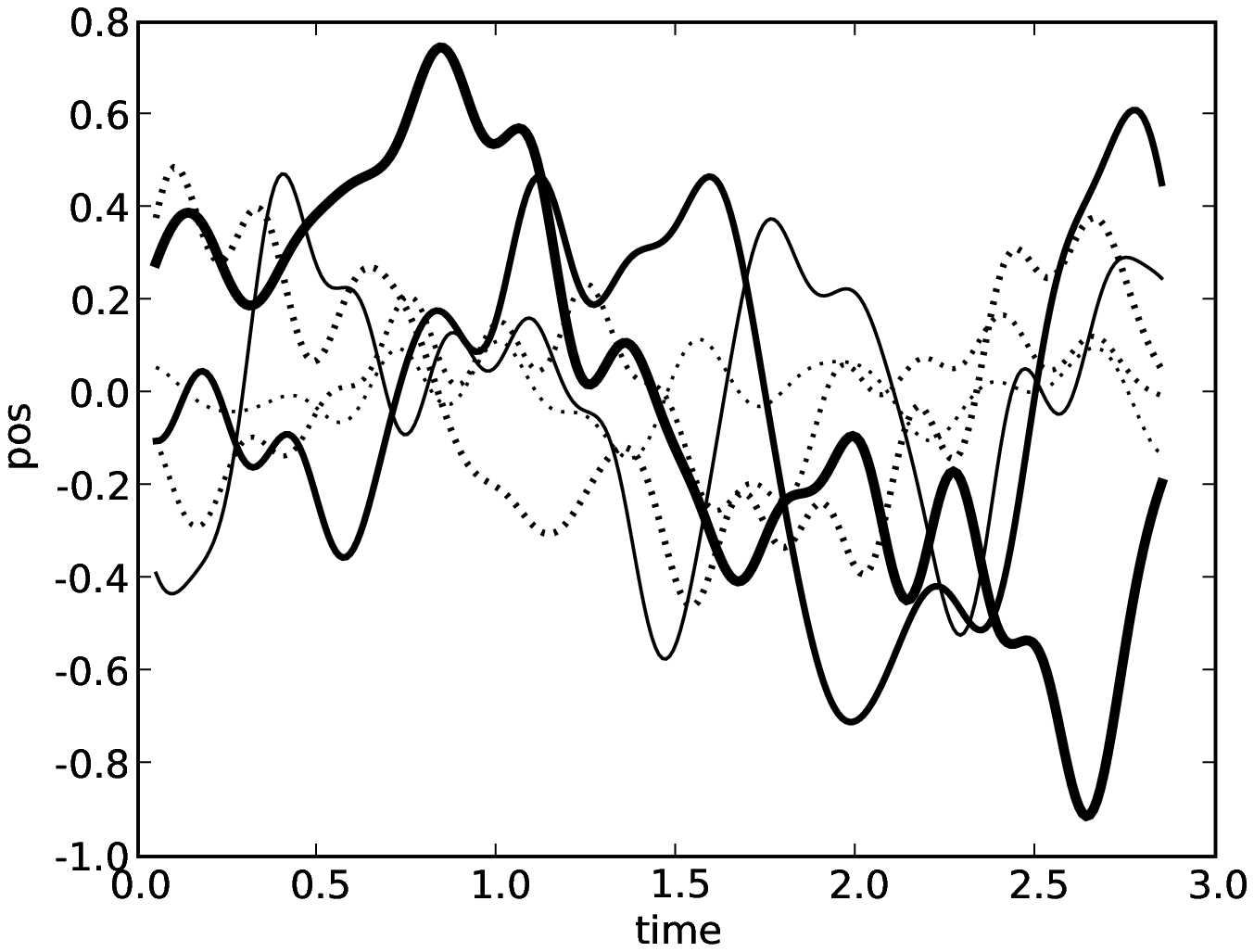}}
    \vskip 4pt
    \hbox to \hsize{\hfill(b)\hfill}}\hfill\strut
  }
  \caption{\label{fig:relax}%
    (a) Diagram of interactions in the HIV capsid lattice.  The black
    and white shapes represent the dimer-forming CTD and the
    hexamer-forming NTD, respectively.  Springs represent the three
    different bonds we are interested in, and dotted lines represent
    the fourth bond we are ignoring.  (b) Relaxation mode trajectories
    of linker.  The mode coordinate has units of $\sqrt{\nano\second}$
    because it has been normalized by the noise.  The slower modes are
    drawn with thicker lines.  Note that the slowest mode has a very
    small drift, and we could correct this by applying an external
    force.  The traces have been smoothed with a low-pass filter for
    readability.}
\end{figure*}

\secc{Application to HIV capsid} The elastic and dynamic properties of
viruses in general are of particular importance in understanding the
mechanisms by which they assemble and disassemble.  The assembly must
be reliable enough to produce capsids capable of surviving the harsh
intercellular environment, while still being able to disassemble upon
entering a new host cell.  HIV in particular is unique because of its
characteristic conical capsids~\cite{li2000irh}, whose mechanism of
formation is yet unsettled.

A capsid is well-modeled by a triangular lattice of proteins, and we
coarse-grain at this level.  We take rigid units to represent either a
whole protein or a sub-domain of a protein.  Each unit therefore
requires six coordinates for its position and orientation.  Provided
the actual interactions are pairwise between units, our program is to
\emph{simulate only a pair of interacting units} at a time, doing a
separate simulation for each kind of unit-unit contact to obtain its
parameters.  The coarse-grained network is then reassembled and
studied using these generalized springs.

The HIV capsid protein (CA) consists of two globular domains: the
larger 145-amino acid N-terminal domain (NTD) has a radius
$1.3\nano\meter$ and the smaller 70-amino acid C-terminal domain (CTD)
has a radius $1.7\nano\meter$; we treat these as two separate units.
The NTD and CTD are connected covalently by a flexible linker; there
is also an NTD--NTD interaction (which forms hexamers in the capsid
structure), a CTD--CTD interaction (which forms symmetric dimers in
the structure), and an NTD--CTD interaction between neighboring
proteins around a hexamer.  These four interactions are shown in
FIG.~\ref{fig:relax}(a).  We believe the NTD--CTD interaction to be
the weakest, and the known structure is also poorest, so we will
ignore it from now on.  We therefore simulate each other pair in
isolation, using structures from the Protein Data
Bank~\cite{bib:structures}.

We carried out our simulations using a modified version of the
NAMD~\cite{bib:MD} package with the CHARMM22 force field.  Our
proteins are in a periodic cell 5 to 9nm to a side using the TIP3P
model for explicit water and 0.1M NaCl, run with 2fs timesteps for a
total of 3ns each.  We do most of the work in at constant pressure and
temperature (NPT), using a Langevin piston barostat at
$P=1\mathrm{atm}$, and a Langevin thermostat at $T=310\kelvin$ and
damping rate $\gamma_\mathrm{L}=5\pico\second^{-1}$.  The NPT
simulations model the statics well, but the thermostat's damping leads
to unphysical dynamics with increased relaxation rates.  This allows
shorter simulations to equilibrate, but prevents us from determining
the rates we should expect to see in reality.  We therefore do a
second measurement of diffusion at constant volume and energy (NVE).

The center of mass and global rotation of the pair accounts for six
trivial degrees of freedom; the remaining six represent the relative
position and orientation of the two domains.  Of these six, only one
is a pure translation: the distance $r=|\vr_2-\vr_1|$ between the
center of each domain.  The orientation of domain $m$ can be
represented by a rotation matrix $\Om_m$ which rotates the domain from
its reference orientation by an angle $|\vth_m|$ about the axis
$\hat\vth_m$.  The even and odd combinations $\vth_1\pm\vth_2$ give
six degrees of freedom that comprise the remaining five coordinates,
along with an overall rotation due to the even combination about the
inter-body axis $\vr_2-\vr_1$.

As we simulate just one pair of units from a protein complex, we omit
the forces and torques on them due to the other units in the lattice,
which generically had a nonzero resultant.  In order to expand the
free energy around the physiologically relevant configuration, we must
add external forces to compensate; in light of \eqref{eq:eom} the
correct force to impose is given by $\vf\eqmin = - \Ga^{-1}\bar\vv$,
where $\bar\vv$ is the drift velocity measured in the absence of the
compensating force.  This was not important for the pairs reported in
our results.

\secc{Results}
The results for each simulation were similar, and the trajectory of
the linker in the transformed relaxation mode coordinates is shown in
FIG.~\ref{fig:relax}(b), which is characteristic of all the observed
trajectories.  Once we have an equilibrated segment of a trajectory we
use~\eqr{eq:Kmeasure} to determine the $6\times 6$ stiffness tensor
$\K$; different components have different units, so it would be
mathematically meaningless to diagonalize it directly.  Instead, we
define reduced stiffness tensors, representing the free energy cost if
we optimize $r$ for a fixed set of angles and vice versa.  Given
\begin{equation}
  \K = \left(\begin{array}{cc} K_{\ur\ur} & \K_{\ur\uth}\\
       \K_{\uth\ur} & \K_{\uth\uth}
        \end{array}\right),
\end{equation}
then
\begin{align}
  K_\str^\eff &= K_{\ur\ur} - \K_{\ur\uth}\K_{\uth\uth}^{-1}\K_{\uth\ur}\\
 \K_\orient^\eff &=  \K_{\uth\uth} - \K_{\uth\ur}K_{\ur\ur}^{-1}\K_{\ur\uth}.
\end{align}
The eigenvalues of the reduced tensors are given in TABLE~\ref{tab:stiff}.

\begin{table}
\begin{tabular}{|l|c|ccccc|}\hline
       & $K_\str^\eff$ & \multicolumn{5}{c|}{$\K_\orient^\eff$ eigenvalues}\\
       &  ($\kB T/\nano\meter\squared$) 
           & \multicolumn{5}{c|} {($\kB T$)} \\
\hline
NTD--NTD   & 12  & 1300 & 2800 & 4500 & 10000 & 18000 \\
CTD--CTD   & 9.9 & 210  & 340  & 1100 & 3900  & 8300  \\
Linker     & 2.8 & 130  & 250  & 480  & 1100  & 3800  \\\hline
\end{tabular}
\caption{\label{tab:stiff}Effective stiffness eigenvalues for pair
  simulations: NTD dimer, CTD dimer, and the NTD--CTD linker within
  the CA protein.}
\end{table}

We computed the stiffness tensor implicitly in the relative
coordinates between the two bodies, but the absolute coordinates are
the natural frame for computing the noise.  Measuring the diffusion of
a single body in an NVE simulation yields a mean
$D\rot\CTD=0.11\radian\squared\per\nano\second$ and
$D\rot\NTD=0.044\radian\squared\per\nano\second$.  If we approximate
each domain as a solid sphere then Stokes' law gives a rotational
diffusion constant $D\rot = \kB T/(8\pi\eta r^3)$~\cite{lamb}.  We
thus expect $D\rot\CTD=0.11\radian\squared\per\nano\second$ and
$D\rot\NTD=0.050\radian\squared\per\nano\second$ using a viscosity
$\eta^\mathrm{(310K)}=0.69\centi\poise$.  The accepted TIP3P viscosity
$\eta^\mathrm{(TIP3P)}=0.31\centi\poise$ gives poorer agreement.

The translational diffusion constant is slightly harder to measure,
since it is influenced significantly by the finite-size
effect~\cite{fse}.  This can be corrected for by measuring the
diffusion at several box side lengths $L$ and using a linear fit of
$D\trans$ versus $1/L$ to extrapolate to $1/L=0$.  Doing so yields
$D\trans\CTD = 55\angstrom\squared\per\nano\second$ and $D\trans\NTD =
27\angstrom\squared\per\nano\second$.  Stokes' law gives expected
$D\trans\CTD=56\angstrom\squared\per\nano\second$ and
$D\trans\NTD=43\angstrom\squared\per\nano\second$ using
$\eta^\mathrm{(TIP3P)}=0.31\centi\poise$.  The measured $D\trans$ has
a significantly larger relative error than $D\rot$, due to the
finite-$L$ extrapolation.

We can diagonalize the relaxation matrix to compute the relaxation
modes for each linkage.  The NPT relaxation times from this
calculation are listed in TABLE~\ref{tab:relaxtimes}.  All the times
are significantly shorter than the simulation time, so we can be
confident that the simulations are equilibrated.

\begin{table}
\begin{tabular}{|l|cccccc|}\hline
       & \multicolumn{6}{c|} {relaxation times $\tau_\alpha$ (ps)} \\
\hline
NTD--NTD   & 120 & 23  & 18 & 9.3 & 6.0 & 4.4  \\
CTD--CTD   & 76  & 26  & 24 & 7.8 & 5.4 & 4.1  \\
Linker     & 190 & 140 & 80 & 76  & 22  & 8.3   \\
\hline
\end{tabular}
\caption{\label{tab:relaxtimes}
NPT time constants for the relaxation modes of each pair.}
\end{table}

Finally, we can compose these generalized springs together into a
triangular lattice as shown in FIG.~\ref{fig:relax}(a), with an NTD
hexamer at each vertex, a CTD dimer at the midpoint of each edge, and
a spring connecting each domain, whose free energy is given by the
relative positions multiplied into the appropriate stiffness tensor.
We can then determine the free energy minimum as a function of
periodic cell dimensions to find a lattice constant of
$a=9.1\nano\meter$.  This is slightly smaller than the experimentally
measured $10.7\nano\meter$~\cite{li2000irh}, which may be largely due
to our sheet being flat, rather than curved into a tube.  Computing
the free energy of simple extension yields a 2d Young's modulus of
$0.92\kB T\per\angstrom\squared=0.39\newton\per\meter$ and a Poisson
ratio of $0.30$.  Assuming homogeneity and a thickness of
$5\nano\meter$, we find a 3d Young's modulus of $77\mega\pascal$
(compared with $115\mega\pascal$ measured using atomic force
microscopy~\cite{kol2007ssh}).

Furthermore, we can estimate the relaxation rate of the full-capsid
breathing mode in water by further coarse-graining to a single
coordinate $a$ representing a uniform dilation in the plane, which has
dynamics given by \eqref{eq:eom} with stiffness and mobility constants
$K_a$ and $\mathit\Gamma_a$.  The projected stiffness is given by the
bulk modulus $K_a=4K_\mathrm{2d}=2.6\kB T\per\angstrom\squared$,
calculated from the 2d Young's modulus and Poisson ratio.  To project
the damping term, we observe that all the actual motion in the
breathing mode of a virus capsid of radius $r$ is in the radial
direction, and we thus need to scale the capsid protein's
translational diffusion constant by $(\ud a/\!\ud r)^2$ to find the
diffusion constant for $a$.  Using the detailed balance condition,
\begin{equation}
  \mathit\Gamma_a = \frac{16\pi\sqrt3}N
                 \frac{D\trans\NTD+D\trans\CTD}{\kB T}
                 \frac{\eta^\mathrm{(TIP3P)}}{\eta^\mathrm{(310K)}},
\end{equation}
where $N=16\pi\sqrt{3}r^2/a^2$ is the total number of capsid
proteins~\cite{note:scale}.  Taking $N=1500$ proteins as the average
size for an HIV capsid thus gives a relaxation rate of
$6.1\nano\second^{-1}$ for the breathing mode.

\secc{Discussion}
In conclusion, we have put forth a model of overdamped random walks in
which the statics and dynamics are described respectively by
complementary ``stiffness'' and ``mobility'' tensors.  From these two
tensors a ``relaxation matrix'' can be formed, the eigenvalues of
which give the relaxation rates, which also provide a convergence test
for simulations.  We demonstrated the usefulness of this model in
extracting coarse-grained elastic constants from molecular dynamics
trajectories of pairs of interacting domains.  HIV is particularly
well-suited for this because the important interactions appear to be
nearest-neighbor, while many other viruses have long tails in which
all six molecules in the hexamer are entwined, making it more
difficult to separate into individual interactions.

Evaluating the forces between protein domains to second order in the
positions and orientations yields a picture of the dynamics that is
simple enough both to simulate with all-atom MD as well as to model at
the coarse-grained level, yet general enough to thoroughly describe
the interaction in the vicinity of the simulated configuration.

Our relaxation formalism bears some similarities to normal mode
analysis, and in particular, Gaussian network models, which replace
atomic interactions by springs of uniform stiffness~\cite{bib:nma}.
While these techniques have been successful in explaining reaction
pathways such as virus maturation~\cite{rader2005mdb,bib:personal},
they suffer from several shortcomings: first, while the normal mode
frequencies are useful in identifying soft degrees of freedom, the
frequencies themselves are well known to be artificial because they
omit the damping forces of the surrounding water.  For instance, the
breathing mode we computed above would have a normal mode frequency of
$\sqrt{K/m} = 60\nano\second^{-1}$.  Additionally, most applications
are coarse-grained to the point that individual residue types are
irrelevant: such a method is entirely insensitive to the effect of
point mutations or of varying the salinity.  Lamm and
Szabo~\cite{lamm1986lmm} introduced so-called ``Langevin modes,''
which are similar to our relaxation modes, but their method still
suffers from the latter issues.

Another quantitative approach to understanding protein dynamics is
``essential dynamics'' (or ``principle component
analysis'')~\cite{bib:ed}.  This technique has the advantage that it
is based on all-atom simulations, with the explicit damping forces and
entropic contributions of the solvent, but the resulting modes can
only be expressed by giving a $3N$-component vector.  Hayward
\textit{et al}~\cite{hayward1997mfm} suggested specifying important
modes \textit{a priori}, and this provides us the great advantage
being able to relate the results of several simulations together into
a bigger picture.  As long as our modes still contain the most
important fluctuations, they are a reasonable basis to use.

We have demonstrated the use of relaxational dynamics in extracting
measurable elastic moduli from small molecular dynamics simulations.
We hope that this technique will provide a convenient middle ground
between the atomistic and continuum pictures for other biological
systems.

\secc{Acknowledgments}
We thank D.~Murray, V.~M.~Vogt, M.~Widom, H.~Weinstein, D.~Roundy,
W.~Sundquist, M.~Yeager, and P.~Freddolino.  This work was supported
by DOE Grant No. DE-FG02-89ER-45405.  Computing facilities were
provided through the Cornell Center for Materials Research under NSF
grant DMR-0079992.


\begin{thebibliography}{16}
\bibitem{note:scale} We have included a term
  $\eta^\mathrm{(TIP3P)}/\eta^\mathrm{(310K)}$ to correct for the
  TIP3P water model's incorrect viscosity (which affected the
  diffusion constants we measured in our simulations) so that we can
  estimate the relaxation rate in real water.
\bibitem{bib:personal}
  E.~R.~May (personal communication) has calibrated the stiffnesses
  in an elastic network model to all-atom MD of the HK97 phage
  (mature) capsid.
\end{thebibliography}
\end{document}